\documentclass[fleqn,usenatbib]{mnras}

\usepackage{newtxtext,newtxmath}
\usepackage[T1]{fontenc}

\DeclareRobustCommand{\VAN}[3]{#2}
\let\VANthebibliography\thebibliography
\def\thebibliography{\DeclareRobustCommand{\VAN}[3]{##3}\VANthebibliography}

\usepackage{graphicx}
\usepackage{amsmath}	% Advanced maths commands

\usepackage{float}
\usepackage{comment}
\usepackage{xcolor}
\usepackage{ulem}
\definecolor{deepblue}{rgb}{0., 0.2, 0.8}
\newcommand{\MM}[1]{\textcolor{black}{#1}}

\newcommand{\MS}[1]{\textcolor{black}{#1}}

\title[The Characteristic Mass and Energy Conversion Efficiency for Cusp–Core Transition]{The Characteristic Mass and Energy Conversion Efficiency in the Cusp–Core Transition of Dark Matter Haloes: Implications for Scaling Relations 
and Supernova feedbacks}

\author[M. Shinozaki et al.]{
 Michi Shinozaki$^1$\thanks{E-mail: shinozaki@ccs.tsukuba.ac.jp},
 Masao Mori$^2$\thanks{E-mail: mmori@ccs.tsukuba.ac.jp},
 Yuka Kaneda$^3$,
 and
 Kohei Hayashi$^{4,5,6}$
\\
$^{1}$Master's Program in Physics, Graduate School of Pure and Applied Sciences, University of Tsukuba, 1-1-1 Tennodai, Tsukuba, Ibaraki 305-8577, Japan\\
$^{2}$Center for Computational Sciences, University of Tsukuba, 1-1-1 Tennodai, Tsukuba, Ibaraki 305-8577, Japan\\
$^{3}$Doctoral Program in Physics, Graduate School of Pure and Applied Sciences, University of Tsukuba, 1-1-1 Tennodai, Tsukuba, Ibaraki 305-8577, Japan\\
$^{4}$National Institute of Technology, Sendai College, 4-16-1 Ayashi-Chuo, Sendai, Japan\\
$^{5}$Astronomical Institute, Tohoku University, 6-3 Aoba, Sendai, Japan\\
$^{6}$ICRR, The University of Tokyo, 5-1-5 Kashiwanoha, Kashiwa, Japan
}

\date{Accepted XXX. Received YYY; in original form ZZZ}

\pubyear{\the\year{}}

\begin{document}
\label{firstpage}
\pagerange{\pageref{firstpage}--\pageref{lastpage}}
\maketitle

\begin{abstract}
Galaxies in the nearby Universe, particularly dwarf systems, exhibit inner mass profiles of dark matter haloes that systematically depart from canonical cold dark matter expectations, signalling an interplay between baryonic feedback and the collisionless halo.
We update an analytical cusp-core transition model by incorporating the effect of supernova-driven mass loss.
Adapting this model to SPARC galaxies, we measure the energy conversion efficiency \(\varepsilon\), defined as the fraction of supernova feedback energy that is used to change the central dark-matter potential.
We find \(\varepsilon \approx 0.01\) for nearby SPARC galaxies. 
Building on these measurements, we compare the dynamical energy required for a cusp-core transformation with the feedback energy available over burst cycles and identify a cusp-core transition forbidden region on the halo-stellar mass plane where transformation cannot occur.
Galaxies with halo masses from \(10^{8}\) to \(10^{11}\,\mathrm{M}_\odot\) lie outside the forbidden region, whereas ultra-faint dwarf galaxies \(<10^{8}\,\mathrm{M}_\odot\), galaxy groups and clusters \(>10^{11}\,\mathrm{M}_\odot\) fall within it, consistent with their high central densities and the inefficiency of core formation at very low and very high masses. This approach also explains the observed diversity of inner density profiles in low-mass systems, showing that both the star formation rate and the energy conversion efficiency govern them, with the latter emerging as a key parameter setting the strength of the cusp-core transition. Beyond the cusp-core problem, our observationally inferred energy conversion efficiency provides a model independent benchmark that strongly constrains galaxy formation models.
\end{abstract}
\begin{keywords}
dark matter --- galaxies: haloes --- galaxies: dwarf --- galaxies: evolution --- galaxies: formation
\end{keywords}

%%%%%%%%%%%%%%%%%%%%%%%%%%%%%%%%%%%%%%%%%%%%%%%%%%

%--------------------------------------------------------------------------------------
%\clearpage
%--------------------------------------------------------------------------------------

%%%%%%%%%%%%%%%%% BODY OF PAPER %%%%%%%%%%%%%%%%%%

%--------------------------------------------------------------------------------------
%
%   Introduction
%
%--------------------------------------------------------------------------------------

\section{Introduction}
The Cold Dark Matter (CDM) model is a cornerstone of modern cosmology, providing successful explanations for the large-scale structure of the Universe, the cosmic microwave background, and the formation of galaxies and galaxy clusters. Despite its predictive power, the model faces challenges. Discrepancies between theoretical predictions and observational evidence have generated significant debate and research. For example, the predicted abundance of subhaloes in the Local Group exceeds observations, which is known as the missing satellites problem, and the inner density profiles of dark matter haloes frequently diverse from theoretical expectations, referred to as the cusp-core problem \citep[e.g.,][]{Moore1994, Burkert1995}.

The cusp-core problem concerns the density profiles of dark matter haloes, particularly in dwarf galaxies and low-surface-brightness galaxies. $N$-body simulations based on the CDM model predict steeply cusped density profiles at the centres of dark matter haloes, represented by a Navarro-Frenk-White (NFW) profile \citep{NavarroFrenkWhite1996}. However, observations often reveal shallower, cored profiles in these regions \citep{Moore1994, Burkert1995}. This discrepancy raises fundamental questions about the nature of dark matter and the processes governing galaxy formation. Early theoretical studies attributed this inconsistency to numerical limitations, such as resolution effects. Yet, as computational techniques advanced, the persistence of the discrepancy suggested that additional physics beyond the standard CDM paradigm might be involved. 

Alternative models to CDM, such as self-interacting dark matter and ultra-light dark matter, have been proposed to address these challenges. These models aim to resolve specific gaps in the CDM paradigm, including the failure to naturally produce cored profiles in low-mass haloes and the suppression of excessive small-scale structures. For example, the lack of cored profiles in low-mass haloes undermines accurate predictions of dwarf galaxy dynamics, while excessive small-scale structures conflict with observed galaxy clustering patterns. 
These models propose physical mechanisms for creating cored profiles \citep{SpergelSteinhardt2000, Hu+2000, Rocha+2013} and have recently attracted attention as promising theoretical frameworks for dark matter.

In contrast, baryonic processes within the CDM framework have been proposed as potential solutions. \citet{ReadGilmore2005} demonstrated that rapid gas expulsion driven by stellar feedback could alter the gravitational potential and induce an outward migration of dark matter particles, thereby forming a central core. Subsequently, supernova-driven feedback has been widely hypothesised to redistribute dark matter in halo centres by generating fluctuations in the gravitational potential \citep{Governato+2010, PontzenGovernato2012}, leading to cusp–core transitions. Extending this line of work with a more rigorous theoretical approach, \citet{OgiyaMori2014} identified Landau resonance as the key mechanism mediating energy transfer between baryonic oscillations and the dark matter potential. This wave–particle interaction facilitates efficient cusp–core transformation and yields a predictive relation between the oscillation period and the resulting core radius.

A series of high-resolution simulations has shown that bursty, time-variable star formation drives the potential fluctuations required to heat dark matter and form cores \citep{Mashchenko+2008, Teyssier+2013, Madau+2014, Chan+2015, Read+2016, Tollet+2016}. Systematic suites of simulations further find that the efficiency of core formation depends strongly on the stellar-to-halo mass ratio \citep{DiCintio+2014, Onorbe+2015}. On the other hand, it is well recognised that outcomes are sensitive to the adopted supernova-feedback model in simulations of galaxy formation, because differences in subgrid parameterisations, coupling schemes, and resolution-dependent calibrations introduce substantial model dependence. Results should therefore, at present,  be evaluated in the context of the chosen implementation and interpreted with caution. Accordingly, it would be desirable to establish an observationally motivated benchmark for galaxy formation that supplies common calibration metrics for the tuning and evaluation of feedback models.

In this context, a central outstanding issue is to identify what controls the efficiency with which feedback couples to the collisionless component. We therefore recast the problem in terms of the energy-conversion efficiency, defined as the fraction of supernova feedback energy converted into work on the gravitational potential of the dark-matter halo. This quantity directly encodes the coupling and provides a target for theoretical prediction and observational constraint. To date there has been no observational investigation of this rate: neither direct determinations from data nor general, model independent measurements exist, and there are no empirical constraints. In this study we develop a framework that links feedback energetics to inner structure, using the energy conversion efficiency as the organising parameter.

From a structural perspective, a number of observational studies have revealed a striking regularity in the surface density of dark matter haloes across a broad range of galaxy types and masses, from dwarfs to spirals \cite[e.g.][]{KormendyFreeman2004,Spano+2008,Donato+2009,Salucci+2012,KormendyFreeman2016}. \citet{Ogiya+2014} demonstrated that the observed scaling relations arise from the concentration-mass ($c$–$M$) relation predicted for CDM haloes, thereby reflecting their fundamental structural properties. Building on this, \citet{KanedaMoriOtaki2024} interpreted this empirical trend within the CDM framework by identifying the radius of maximum circular velocity, $R_{\mathrm{max}}$, as a physically meaningful scale. At this radius, the characteristic mean surface density, \MS{$\bar\Sigma_{R_{\mathrm{max}}}$}, as defined by \citet{Hayashi+2017}, shows remarkable agreement between observations and predictions based on the $c$–$M$ relation derived from cosmological simulations.

Guided by these considerations, we organise the next steps around two complementary elements, a structural diagnostic and an energetic criterion. First, we adopt $\bar\Sigma_{R_{\mathrm{max}}}$ as a robust observable that links nearby galaxy data to the mass to concentration relation, thereby setting the baseline CDM expectation. Second, we frame feedback coupling through the energy conversion efficiency, defined as the fraction of supernova energy that does work on the central dark matter potential. This energetic view allows a direct comparison between the dynamical work required to flatten a cusp and the feedback energy supplied over burst cycles, and it predicts a mass dependent transition in which coupling is most effective near a characteristic halo mass.

This paper elucidates the physical basis for these features. By combining analytical estimates, calculations using state-of-the-art cosmological $N$-body simulations \citep{Ishiyama+2021}, and constraints from the kinematics of neighbouring galaxies~\citep{Hayashi+2025}, we construct a unified framework that links feedback energetics to inner halo structure. Within this framework we investigate whether a characteristic mass scale emerges that separates haloes whose inner profiles tend to remain cusped from those susceptible to core formation, and we examine the underlying physics that governs this distinction. We also derive the energy conversion efficiency directly from observations using SPARC data, and test whether it acts as a primary control parameter for inner halo structure and for the observed diversity of profiles.

This paper is organised as follows. Section~2 introduces the theoretical model and derives the conditions for transformation in terms of the feedback energy budget and the energy conversion efficiency. Section~3 confronts the model with observational data and simulation-based inferences, including constraints on the energy conversion efficiency for nearby galaxies. Section~4 discusses the implications and limitations of our results. Finally, Section~5 concludes.

%-----------
% Fig. 1
%-----------
\begin{figure*}
    \centering
    \includegraphics[width=1.0\linewidth]{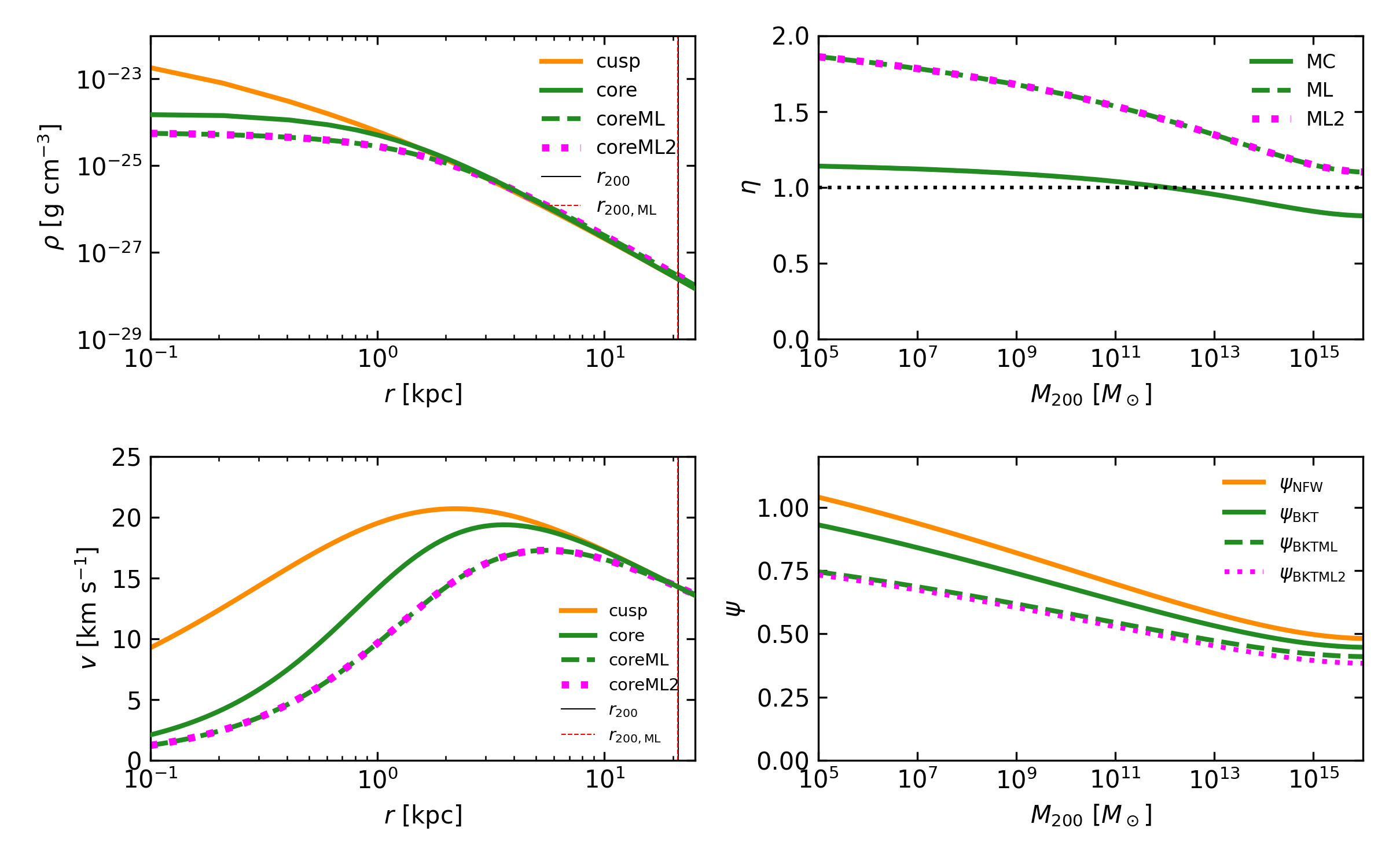}
    \caption{
    \textbf{Left:} Radial profiles of density (upper-left) and circular velocity (lower-left) for a dark matter halo with mass $M_{200} = 10^9\,M_{\odot}$, illustrating the effect of the cusp--core transition. The orange curve represents the NFW profile. 
    The solid and dashed green curves indicate Burkert profiles with mass loss fractions $f = 0$ and $f = f_\mathrm{b} $, respectively. 
    The magenta dashed curve indicates the model where the virial radius is recalculated from the halo mass after mass loss, adopting a mass–loss function of $f = f_{\mathrm{b}}$.
    \textbf{Upper-right:} The dependence of $\eta$ on $M_{200}$. Solid and dashed curves correspond to mass loss fractions of $f = 0$ and $f = f_\mathrm{b} $, respectively.  
    \textbf{Lower-right:} The dependence of $\psi_{\mathrm{NFW}}$ and $\psi_{\mathrm{BKT}}$ on $M_{200}$. The orange curve represents the NFW profile, while 
    the solid and dashed green curves denote Burkert profiles with $f = 0$ and $f = f_\mathrm{b} $, respectively.
    The magenta dashed curve indicates a marginally smaller $\psi$ resulting from the redefinition of the virial radius.
    }  
    \label{fig:fig1}
\end{figure*}
%-----------

\section{Theoretical model} \label{sec:theoreticalmodel}

We adopt the premise that dark matter haloes initially formed with cuspy density profiles, which are later transformed into cored profiles through subsequent dynamical interactions. 
To capture this transformation, we develop a simplified analytical framework that links the parameters describing the initial cuspy profile, namely the scale density and scale radius, to those characterizing the resultant cored profile, namely the central density and the core radius.

In an earlier analytical study of the cusp-core transition, \citet{Ogiya+2014} formulated a dynamical mechanism under the approximation that the transition region is small compared with the halo scale length, whereas \citet{KanedaMoriOtaki2024} relaxed this restriction and developed a more general, self-consistent model. Here we extend that framework by allowing mass loss during the transition, yielding a more realistic description that underpins the analyses below. We then introduce an energy transport model that quantifies how feedback energy couples to the central potential, derive the critical stellar mass required to supply the transition energy, and finally embed these relations in a three-dimensional parameter space linking halo mass, baryonic content, and structural diagnostics.

\subsection{Cusp-core transition model}

In this paper, we adopt the NFW density profile
\begin{equation}
  \rho_\mathrm{NFW}(r)
  = \frac{\rho_{\mathrm N}}{\left(r/r_{\mathrm N}\right)\left(1+r/r_{\mathrm N}\right)^{2}},
\end{equation}
to represent cusped haloes, and the Burkert profile
\begin{equation}
  \rho_\mathrm{BKT}(r)
  = \frac{\rho_{\mathrm B}}{\left(1+r/r_{\mathrm B}\right)\bigl[1+\left(r/r_{\mathrm B}\right)^{2}\bigr]},
\end{equation}
to represent cored haloes. Here, \(\rho_{\mathrm N}\) and \(r_{\mathrm N}\) denote the NFW scale density and scale radius, and \(\rho_{\mathrm B}\) and \(r_{\mathrm B}\) denote the Burkert core density and core radius; throughout, the subscripts \(\mathrm N\) and \(\mathrm B\) indicate quantities defined for the NFW and Burkert profiles, respectively. Both profiles share the same asymptotic behaviour, \(\rho \propto r^{-3}\) as \(r \to \infty\), which facilitates a simple, self-consistent model.

In \citet{Ogiya+2014} and \citet{KanedaMoriOtaki2024}, the cusp-core transition model was constructed under the assumption of mass conservation before and after the transition. In contrast, the present study develops a model that incorporates mass loss from the system due to Type II supernova (SNII) feedback.
One key assumption in our model is that the change in density distribution during the transition occurs only in the central region of the halo, while the density around the virial radius remains unchanged before and after the cusp–core transition.

The condition implied by this assumption can be written as

\begin{equation}
\frac{\rho_\mathrm{N} }{\frac{r_{200}}{r_\mathrm{N} }\left(\frac{r_{200}}{r_\mathrm{N} }+1\right)^2} = \frac{\rho_\mathrm{B} }{\left(\frac{r_{200}}{r_\mathrm{B} }+1\right)\left[\left(\frac{r_{200}}{r_\mathrm{B} }\right)^2+1\right]},
\end{equation}
where $r_{200}$ defined as the radius within which the mean density is 200 times the critical density of the Universe and the relation between the virial masses before and after the transition, incorporating the fractional mass loss $f$, is given by
\begin{equation}
(1-f) \rho_\mathrm{N}  r_\mathrm{N} ^3 F_\mathrm{N} \left(\frac{r_{200}}{r_\mathrm{N} }\right) = \rho_\mathrm{B}  r_\mathrm{B} ^3 F_\mathrm{B} \left(\frac{r_{200}}{r_\mathrm{B} }\right),
\end{equation}
where the functions $F_\mathrm{N} (x)$ and $F_\mathrm{B} (x)$ are defined as
\begin{align}
F_\mathrm{N} (x) &= \ln(1+x) - \frac{x}{1+x},\\
F_\mathrm{B} (x) &= -\frac{1}{2}\arctan(x) + \frac{1}{2}\ln(1+x) + \frac{1}{4}\ln(1+x^2).
\end{align}

The mass loss fraction $f$ varies between 0 and $f_\mathrm{b}$, where $f_\mathrm{b}$ represents the cosmological baryon fraction. Based on cosmological parameters, this is given by $f_\mathrm{b}  = \Omega_\mathrm{b}  / \Omega_m \approx 0.1565$, where $\Omega_\mathrm{b} $ and $\Omega_m$ are the baryon and matter density parameters, respectively. 

According to the Planck 2018 results \citep{Planck2020}, we adopt $\Omega_{\mathrm b}h^{2}=0.0224, \, h=0.674, \, \Omega_{m}=0.315$.

Introducing $\eta = r_\mathrm{B} /r_\mathrm{N} $ and the concentration parameter $c_{200} = r_{200}/r_\mathrm{N} $, we obtain the following equation:
\begin{equation}
g(\eta) - (1-f) \frac{F_\mathrm{N} (c_{200})}{F_\mathrm{B} (c_{200}/\eta)} = 0,
\end{equation}
with
\begin{equation}
g(\eta) \equiv \frac{(c_{200}+\eta)(c_{200}^2+\eta^2)}{c_{200}(c_{200}+1)^2}.
\end{equation}

The concentration parameter $c_{200}$ is known to be a function of $M_{200}$ through the $c$--$M$ relation.
Here, we adopt the $c$–$M$ relation provided in \citet{Moline2023} and \citet{KanedaMoriOtaki2024}, which is for subhaloes in mass range $10^7\sim10^{15}~M_\odot$ based on Uchuu simulation \citep{Ishiyama+2021},
given by
\begin{equation}
c_{200} = \frac{r_{200}}{r_\mathrm{N} } = c_0 \left\{ 1+ \sum_{i=1}^3 a_i \left[\log_{10} \left(\frac{M_{200}}{M_{\odot}}\right)\right]^i \right\}, 
\label{eq:c200_definition}
\end{equation}
with
\begin{equation}
c_0 = 48.29,\quad a_i = [-0.01863,\ -0.008262,\ 0.0003651].
\end{equation}
Therefore, $\eta$ is uniquely determined as a function of $M_{200}$ via \MS{equation~\eqref{eq:c200_definition}}, once the pre-transition halo mass $M_{\mathrm{NFW}}(r_{200}) \equiv M_{200}$ is specified.

In what follows we consider two cases: (i) $f=0$ (mass conservation), and (ii) $f=f_{\mathrm b}$ (maximal baryonic mass loss, hereafter referred to as mass-loss
or ML case). 
In the latter case, we also consider the case with the change in the virial radius due to mass loss is also taken into account (hereafter referred to as mass-loss~2 or ML2 case)

\begin{equation}
r_{200, ML} = \left(\frac{3}{4} \frac{(1-f)M_{200}}{200\pi \rho_{crit,0}}\right)^{\frac{1}{3}}. 
\end{equation}

The resulting changes in the density profile and rotation curve for a halo with $M_{200} = 10^9 M_{\odot}$ that initially follow the $c$--$M$ relation are shown in the upper-left and lower-left panels of Fig.~\ref{fig:fig1}. With increasing $f$, the post-transition maximum circular velocity $V_{\mathrm{max}}$ decreases, and its radius $r_{\mathrm{max}}$ increases. The mass-loss and mass-loss 2 cases differ only in the definition of the virial radius, while their internal structures are identical. Consequently, the upper-left and lower-left panels of Fig. ~\ref{fig:fig1} show no discernible difference between the two models. The upper-right panel of Fig.~\ref{fig:fig1} also shows how $\eta$ varies for the cases of $f = 0$ and $f = f_\mathrm{b}$. Notably, while $\eta \sim 1$ in the mass-conserving (MC) case, we find $\eta > 1$ across eleven orders of magnitude in $M_{200}$ when $f = f_\mathrm{b}$. Furthermore, since the mass-loss and mass-loss 2 cases share the same internal structure—namely identical values of $r_{\mathrm{S}}$ and $r_{\mathrm{B}}$—they naturally have the same value of $\eta$.

%-----------
% Fig. 2
%----------- 
\begin{figure*}
	\includegraphics[width=1.0\linewidth]{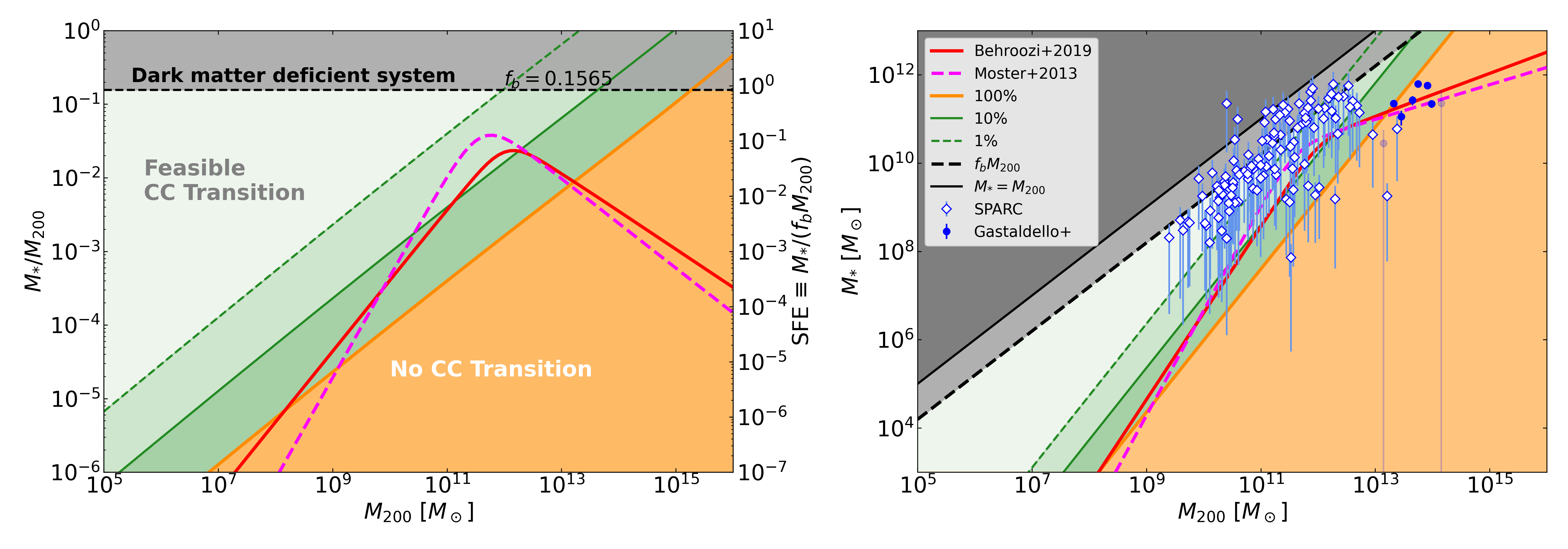}
    \caption{
    The left panel shows $M_{\ast,\mathrm{crit}}/M_{200}$ versus $M_{200}$; the right panel shows $M_{\ast,\mathrm{crit}}$ versus $M_{200}$. The orange solid, green solid, and green dashed lines correspond to $\varepsilon=1$, $0.1$, and $0.01$, respectively. The red solid and magenta dashed curves show the $z=0$ stellar-to-halo mass relations from \citet{Behroozi+2019} and \citet{Moster+2013}, respectively. Regions with star formation efficiency (SFE) $>1$ are shaded gray. In the left panel, open white symbols indicate SPARC data, and filled blue symbols denote the sample of \citet{Gastaldello+2007}. Objects in \citet{Gastaldello+2007} with stellar-mass uncertainties larger than their nominal values are shown in light blue.
    }
    \label{fig:fig2}
\end{figure*}
%-----------

\subsection{Energy transport model}

In this section, we quantify the change in the total energy of a halo before and after the cusp–core transition. We define the transition energy $\Delta E$ as the difference in total energy:
\begin{align}
\Delta E &\equiv E_{\mathrm{BKT}} - E_{\mathrm{NFW}}, \\
&= -(\psi_{\mathrm{BKT}} - \psi_{\mathrm{NFW}}) \frac{GM_{200}^2}{r_{200}}, \\
&= -(\psi_{\mathrm{BKT}} - \psi_{\mathrm{NFW}}) M_{200} \phi_{200},
\end{align}
where the gravitational potential at the virial radius is defined as
\begin{equation}
\phi_{200} = \frac{GM_{200}}{r_{200}}.
\end{equation}

For convenience, the pre- and post-transition energies, $E_{\mathrm{NFW}}$ and $E_{\mathrm{BKT}}$, are expressed using dimensionless correction factors, $\psi_{\mathrm{NFW}}$ and $\psi_{\mathrm{BKT}}$ (functions of $M_{200}$), which measure the deviation from the point-mass scaling $G M_{200}^{2}/r_{200}$. The NFW coefficient is obtained analytically,
\begin{align}
\psi_{\mathrm{NFW}}(c_{200}) =
\frac{c_{200}\!\left[c_{200}(2+c_{200}) - 2(1+c_{200})\ln(1+c_{200})\right]}
{4\,\left[(1+c_{200})\ln(1+c_{200}) - c_{200}\right]^{2}}.
\end{align}
By contrast, no closed-form expression is available for the Burkert profile; we therefore compute it numerically and represent it with the following fitting function:
\begin{align}
\log_{10}\left[\psi_{\mathrm{BKT}}(M_{200}/M_{\odot})\right] &= \sum_{i=0}^{3} a_i 
\left[ \log_{10}\left( \frac{M_{200}}{M_{\odot}} \right) \right]^i \, 
\end{align}
The best-fitting coefficients are
$a_0 = -0.1121$, $a_1 = 0.04690$, $a_2 = -0.007454$, and $a_3 = 0.0002226$.
The maximum relative deviation between the numerical results and the fitting function is less than $0.642\%$ within the mass range where the $c$--$M$ relation is considered reliable \ ($8.0<\log_{10}(M_{200}/M_{\odot})<13.0$), indicating that the approximation is sufficiently accurate for our analysis.

The lower-right panel of Fig.~\ref{fig:fig1} shows both $\psi_{\mathrm{NFW}}$ and $\psi_{\mathrm{BKT}}$. Since $\psi_{\mathrm{NFW}}$ is common to both cases $f = 0$ and $f = f_\mathrm{B} $, the two different values of $\psi_{\mathrm{BKT}}$ are shown as green and light green curves, respectively. The difference between these two curves directly corresponds to the transition energy $\Delta E$. It is found that the energy required for the transition is typically about 10\% of the pre-transition total energy, i.e. $\Delta E/E_{\mathrm{NFW}} \sim 0.1$.
Comparing the mass-loss and mass-loss2 cases, the latter shows a slightly smaller $\psi_{\mathrm{BKT}}$ due to the inward shift of the virial radius caused by mass loss. However, since this difference is minor, we consider only the mass-loss case as the representative model in the following discussion.

We define the energy conversion efficiency $\varepsilon$ as the fraction of SNII energy that is available to drive the cusp–core transition. When the total energy released by SNII balances the transition energy, the following relation holds:
\begin{align}
\Delta E 
&= \varepsilon E_{\mathrm{SNII}}, \\ \label{eq:energytransfer}
&= \varepsilon\, e_{51} f_{\mathrm{SNII}} M_{\ast,\mathrm{crit}},
\end{align}
where $M_{\ast,\mathrm{crit}}$ denotes the critical stellar mass — the minimum stellar mass required to induce the cusp–core transition via SNII feedback.

We now estimate the energy input from SNII,
\begin{align}
E_{\mathrm{SNII}} = e_{51}\, f_{\mathrm{SNII}}\, M_{\ast},
\end{align}
where $e_{51}=10^{51}\ \mathrm{erg}$ is the energy per SNII event, $M_{\ast}$ is the stellar mass, and $f_{\mathrm{SNII}}$ is the number of SNII per unit stellar mass formed. Assuming all stars with initial mass $m>8\,M_{\odot}$ explode as SNII, we compute
\begin{align}
f_{\mathrm{SNII}} = \int_{8\,M_{\odot}}^{100\,M_{\odot}} \phi(m)\,\mathrm{d}m,
\end{align}
with the initial mass function~(IMF) $\phi(m)$ normalised as defined in the following equation.
\begin{equation}
    1 = \int_{0.1\,M_{\odot}}^{100\,M_{\odot}} m \phi(m)dm
\end{equation}

Here, we consider three representative IMFs (Salpeter \citep{Salpeter1955}, Kroupa \citep{Kroupa2001}, and Chabrier \citep{Chabrier2005}). The resulting values of \(f_{\mathrm{SNII}}\) are
\(7.4217\times10^{-3}\) (Salpeter),
\(12.846\times10^{-3}\) (Kroupa),
and \(14.335\times10^{-3}\) (Chabrier).
This fraction is larger for the Kroupa and Chabrier IMFs than for the Salpeter IMF. The difference arises from their low-mass behaviour below \(1\,M_\odot\): whereas Salpeter continues as a single power law toward lower masses, Kroupa flattens and Chabrier turns over, reducing the relative number of low-mass, non-SNII projenitors and thereby increasing \(f_{\mathrm{SNII}}\).
In this study, we adopt the Chabrier IMF, which yields the largest \(f_{\mathrm{SNII}}\) and thus maximises the available SN II energy per unit stellar mass, a conservative choice when assessing the critical condition.

\subsection{Critical stellar mass}

%-----------
% Fig. 3
%-----------
\begin{figure*}
    \centering
    \includegraphics[width=1.0\linewidth]{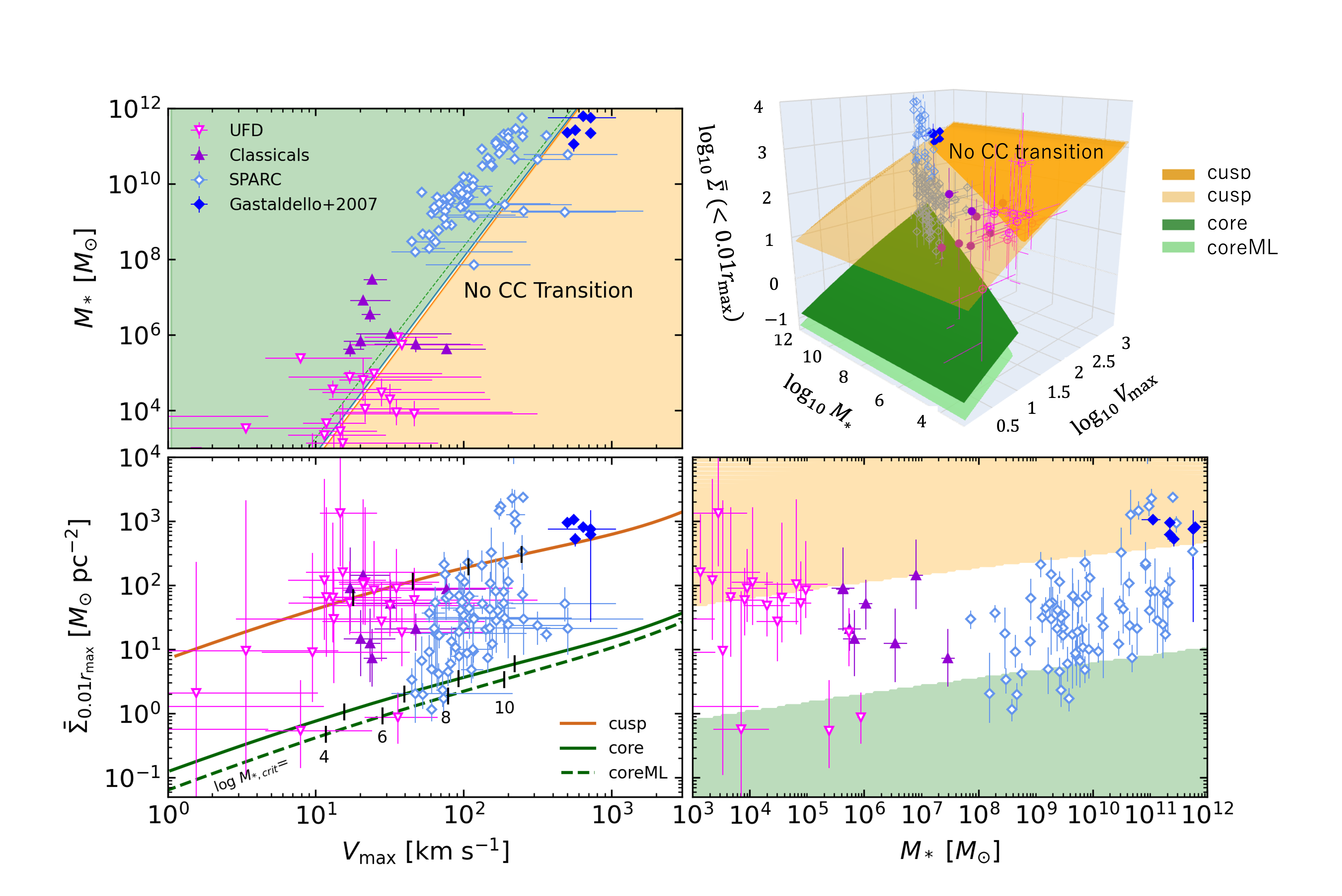}
    \caption{
    \textbf{Upper-right:} The full light orange plane represents the surface in the three-dimensional parameter space followed by NFW dark matter haloes determined by the $c$–$M$ relation. The dark orange region indicates the forbidden zone for the cusp-core transition, derived using our energy transport model explained in section \ref{sec:theoreticalmodel} assuming energy conversion efficiency $\varepsilon = 1$. The planes followed by Burkert-profile haloes, whose parameters derived by our cusp-core transition model explained in section \ref{sec:theoreticalmodel}, are shown in dark green and light green for cases the mass conservation model and mass loss model, respectively.  
    \textbf{Upper-left:} The $V_{\mathrm{max}}$–$M_{\ast}$ projection, corresponding to a top-down view of the upper-right panel. 
    The green solid and dashed lines show $M_{*, \mathrm{crit}}$ as a function of $V_{\rm{max}}$ using for mass conservation model and mass loss model, respectively. Here, we assume $\varepsilon=1$. The orange solid line also shows $M_{*, \mathrm{crit}}$ as a function of $V_{\rm{max}}$, but $V_{\rm{max}}$ is calculated for the cusp profile.
    \textbf{Lower-left:} The $V_{\mathrm{max}}$–$\bar{\Sigma}(<0.01\,r_{\mathrm{max}})$ projection, corresponding to the upper-right panel seen from lower-right.
    The orange solid curve shows $c$--$M$ relation converted on the $V_{\mathrm{max}}$–$\bar{\Sigma}(<0.01\,r_{\mathrm{max}})$ plane.
    The solid and dashed green curves show the $c$--$M$ relation for cored haloes, which is derived using our cusp-core transition model for the mass conservation case and the mass loss case, respectively. Here, we assume $\varepsilon=1$ again.
    Tick marks on green curves for $V_{\mathrm{max}}$ correspond to $M_{\ast,\mathrm{crit}}/M_{\odot} = 10^4,\ 10^6,\ 10^8$ and $\ 10^{10}$.  
    \textbf{Lower-right:} The $M_{\ast}$–$\bar{\Sigma}(<0.01\,r_{\mathrm{max}})$ projection, corresponding to the upper-right panel seen from lower-left. In both the upper-left and lower-right panels, the cusp region shown in orange includes only the forbidden region from the upper-right panel.
    }
    \label{fig:fig3}
\end{figure*}

Solving equation~\eqref{eq:energytransfer} for $M_{\ast,\mathrm{crit}}$, we obtain:
\begin{align} \label{M_star_crit}
M_{\ast,\mathrm{crit}}
&= \frac{\Delta E}{\varepsilon \, e_{51} f_{\mathrm{SNII}}}, \\
&= (\psi_\mathrm{NFW} - \psi_\mathrm{BKT}) \frac{ M_{200} \phi_{200} } {\varepsilon \, e_{51} f_{\mathrm{SNII}}}.
\end{align}

Any galaxy with stellar mass below $M_{\ast,\mathrm{crit}}$ is, in principle, incapable of inducing a cusp–core transition via SNII feedback. This defines the orange-shaded "forbidden region" in Fig.~\ref{fig:fig2}, corresponding to haloes where such a transition cannot occur. The figure is based on $\Delta E$ calculated for the $f = 0$ (mass-conserving) case.

In Fig.~\ref{fig:fig2}, the ratios $M_{\ast,\mathrm{crit}}/M_{200}$ and the corresponding $M_{\ast,\mathrm{crit}}$ values are plotted for energy conversion efficiencies $\varepsilon = 1$, 0.1, and 0.01, using orange, green, and green dashed lines, respectively. The star formation efficiency (SFE) is defined such that SFE = 1 when $M_{\ast}/M_{200} = f_\mathrm{b}$.

The lower-left panel of Fig.~4 of \citet{KanedaMoriOtaki2024}, which compares theoretical prediction and observation on characteristic mean surface density within $0.01r_{\mathrm{max}}$ 
\MS{
, where $r_{\mathrm{max}}$ is the radius at the maximum circular velocity, 
}
against $M_{200}$. 

Here, the characteristic mean surface density is defined by

\begin{equation}
    \bar{\Sigma}(<r) = \frac{M(r)}{\pi r^2},
\end{equation}
where
\begin{equation}
    M\left(r\right)=\int_0^{r} 4 \pi \rho_{\mathrm{dm}}\left(r^{\prime}\right) r^{\prime 2} d r^{\prime},
    \label{eq:mass_definition}
\end{equation}
% shinozaki 
\MS{
where $\rho_{\mathrm{dm}}$ corresponds to the density profile of a dark matter halo~\citep[see also][]{Hayashi+2015,Hayashi+2017}.
}
\MS{
It should be noted that equation~\eqref{eq:mass_definition} is defined as the three-dimensional mass enclosed within a sphere divided by the projected area, rather than the mass contained within a cylindrical volume along the line of sight.
}

This parameter is proposed 
1) to compare the dark matter haloes that have various profiles, as one can specify any profile for $\rho_{\mathrm{dm}}$
2) to compare the dark matter haloes that are associated with a wide dynamic range of galaxies from dwarf galaxies to clusters of galaxies, as $\bar{\Sigma}(<0.01r_{\mathrm{max}})$ adjust the radius to take the mean depending on the scale of dark matter haloes represented by $r_{\mathrm{max}}$.
As seen in the lower-left panel of Fig.~4 of \citet{KanedaMoriOtaki2024}, low-mass dark-matter haloes are consistent with cored profiles, whereas massive haloes favour cuspy profiles for $M_{200}\gtrsim 10^{11}\,M_{\odot}$.

Accordingly, we define the mass range where we can see the change in trend from the cuspy profile to the cored profile as $5.0 \times 10^{10} \lesssim M_{200}/M_{\odot} \lesssim 5.0 \times 10^{11}$.
From the Fig.~\ref{fig:fig2}, we find that galaxies within this transition mass range generally exhibit high SFE and lie well above the $M_{\ast,\mathrm{crit}}/M_{200}$ threshold, making SNII-driven cusp–core transitions feasible. In contrast, galaxy groups with larger halo masses tend to fall below this threshold, indicating that such transitions are not energetically viable in these systems. Therefore, within the CDM framework, such haloes are expected to retain cusps.

In Fig.~\ref{fig:fig2}, the red solid and the magenta dashed curves show the $z=0$ stellar–halo mass relations of \citet{Behroozi+2019} and \citet{Moster+2013}, respectively. From these relations, we infer that, under the \citet{Behroozi+2019} track, the energy conversion efficiency $\varepsilon$ falls below $0.1$ for virial masses in the range $10^{11}$–$10^{12}\,M_{\odot}$, while under the \citet{Moster+2013} track it falls below $0.1$ across $10^{10}$–$10^{12}\,M_{\odot}$. In other words, galaxies within these mass intervals are comparatively prone to undergo a cusp-core transition, since only several per cent of the available supernova feedback energy is required.

It is important to note that $M_{\ast,\mathrm{crit}}$ as given by equation~\eqref{M_star_crit} refers to the stellar mass at the time of star formation, i.e. before any supernovae have occurred. Therefore, when comparing with observed galaxies—where SNII have already taken place—$M_{\ast,\mathrm{crit}}$ may therefore overestimate the present stellar mass.
For example, to account for this, we estimate the stellar mass fraction contributed by stars with $m < 8\,M_{\odot}$ that do not undergo SNII, using the Chabrier IMF \citep{Chabrier2005}. The resulting mass fraction is
\begin{equation}
f_\mathrm{M} = \int_{0.1\,M_{\odot}}^{8\,M_{\odot}} m\, \phi(m)\, \mathrm{d}m \approx 0.732.
\end{equation}

Note that this $f_\mathrm{M}$ includes only main-sequence stars and does not account for the mass locked in remnants such as neutron stars or black holes. This factor should be considered when comparing with observational data.

\subsection{Three-dimensional parameter space}

In this section, we discuss the constraints on the cusp--core transition driven by SNII feedback energy, as derived from the critical stellar mass $M_{\ast, \mathrm{crit}}$, within a three-dimensional parameter space. These constraints are combined with the $c$--$M$ relation obtained in previous studies.

The three parameters considered are as follows:
\begin{itemize}
    \item Virial mass of the halo, $M_{200}\ [M_{\odot}]$, 
    \item Characteristic mean surface density within $r<0.01\,r_{\mathrm{max}}$, $\bar{\Sigma}(<0.01\,r_{\mathrm{max}})\ [M_{\odot}\,\mathrm{pc}^{-2}]$, 
    \item Stellar mass, $M_{\ast}\ [M_{\odot}]$.
\end{itemize}

To compare with observational data, we convert the virial mass $M_{200}$ to the maximum circular velocity $V_{\max}\,[\mathrm{km\,s}^{-1}]$, which is directly observable. For an NFW halo,
\begin{align}
    V_{\max} &= \left[\frac{G\,M(r_{\max})}{r_{\max}}\right]^{1/2}, \label{Vmax_def}\\
             &= \left[\frac{F_{\mathrm N}(x_{\mathrm N})}{F_{\mathrm N}(c_{200})}\,\frac{c_{200}}{x_{\mathrm N}}\,\phi_{200}\right]^{1/2},
    \label{Vmax}
\end{align}
where $x_{\mathrm N}\equiv r_{\max}/r_{\mathrm N}=2.1626$. For a Burkert halo, replace $F_{\mathrm N}\to F_{\mathrm B}$ and $x_{\mathrm N}\to x_{\mathrm B}$, where $x_{\mathrm B}\equiv r_{\max}/r_{\mathrm B}=3.2446$.

In the upper-right panel of Fig.~\ref{fig:fig3}, the semi-transparent orange surface on the near side marks the region where a cusp-core transition is possible, whereas the opaque orange surface on the far side marks the transition-forbidden region (for $\varepsilon=1$, owing to insufficient SNII feedback energy); the two surfaces are connected along the locus of $M_{\ast,\mathrm{crit}}$, as also shown in Fig.~\ref{fig:fig2}. The green surface and the underlying light-green surface correspond to Burkert haloes with $f=0$ and $f=f_{\mathrm b}$, respectively.
Within the standard $\Lambda$CDM framework, dark-matter haloes are initially expected to populate the orange surface. Only those that lie outside the transition-forbidden region and have sufficient energy can undergo the cusp-core transition, reducing the central density and moving from the cusp surface to the core surface.

The upper-left, lower-left, and lower-right panels of Fig.~\ref{fig:fig3} show the projections onto the $V_{\max}$–$M_{\ast}$, $V_{\max}$–$\bar{\Sigma}(<0.01\,r_{\max})$, and $M_{\ast}$–$\bar{\Sigma}(<0.01\,r_{\max})$ planes, respectively.
In the $V_{\max}$–$M_{\ast}$ plane (upper-left panel), the loci of $M_{\ast,\mathrm{crit}}$ are offset, indicating differences between the mass-conserving (green solid) and mass-loss (green dashed) models. The orange solid line shows $M_{\ast,\mathrm{crit}}$ versus $V_{\max}$, with $V_{\max}$ computed for the cusp profile. Similarly, in the $V_{\max}$–$\bar{\Sigma}(<0.01\,r_{\max})$ plane (lower-left panel), the orange solid curve shows $\bar{\Sigma}(<0.01\,r_{\max})$ for the cusp profile; the green solid and green dashed curves denote the mass-conserving and mass-loss models, respectively.
The $M_{\ast}$–$\bar{\Sigma}(<0.01\,r_{\max})$ projection (lower-right panel) corresponds to the upper-right panel viewed from the lower-left. The orange region indicates the cusp region, and the green region indicates the core region.

\section{Comparison with observational data}
In this section we interpret the observational data using our new cusp-core transition model with mass loss. 
We first outline the data sets and the mapping from measured kinematics to $M_{200}$, $V_{\max}$, and the characteristic mean surface density $\bar{\Sigma}(<0.01\,r_{\max})$ within $0.01\,r_{\max}$; we then place the galaxies in the three-dimensional parameter space and quantify the observed scatter relative to the model tracks. Finally, we infer the energy conversion efficiency $\varepsilon$ from the data.
\MM{
On this basis, we extend the interpretation of $\varepsilon$ beyond its simple definition as the fraction of the supernova feedback energy available to alter the gravitational potential of the dark matter halo. Assuming that a cusp-core transition has occurred, we estimate $\varepsilon$ using stellar masses and halo properties inferred from observations. In this context, $\varepsilon$ denotes the effective energy conversion efficiency required to produce a fully cored profile.
}

\subsection{Details of the Observational Data}

\textit{UFDs (Ultra Faint Dwarfs):}

We adopt the data presented in \citet{Hayashi+2023}. In their study, axisymmetric Jeans equations are solved for 25 ultra-faint dwarf galaxies, assuming an axisymmetrically generalised Hernquist profile for the dark matter density, and an axisymmetric Plummer model for the stellar density profile. The maximum circular velocity, $V_{\mathrm{max}}$, is computed from the best-fitting dark matter and stellar profile parameters, obtained via an MCMC analysis. The corresponding $r_{\mathrm{max}}$ is then derived, enabling computation of $\bar{\Sigma}(<0.01\,r_{\mathrm{max}})$. 
Stellar masses are estimated from observational compilations \citep{Bechtol+2015,McConnachie2012,Ji+2021}.

\textit{Classical dwarf spheroidal galaxies:}

We employ the data from \citet{Hayashi+2020}, who also solve axisymmetric Jeans equations under the same assumptions: a generalised Hernquist profile for the dark matter and an axisymmetric Plummer model for the stellar distribution. The parameter estimation methodology is identical to that used for the UFDs.
Stellar masses for classical galaxies are estimated in the same manner as for UFDs.

\textit{SPARC galaxies (massive galaxies):}

We use the results from \citet{Hayashi+2025}, who estimated the dark matter density profiles of galaxies by assuming an axisymmetrically generalised Hernquist profile, utilizing rotation curve data from the SPARC (Spitzer Photometry \& Accurate Rotation Curves) galaxy catalogue\footnote{\url{http://astroweb.cwru.edu/SPARC/}}.
When estimating the supernova feedback energy, we use the total baryonic mass of each galaxy.

\textit{Gastaldello sample (galaxy groups):}

\citet{Gastaldello+2007} analyse 16 relaxed galaxy groups and poor clusters selected from Chandra and XMM-Newton X-ray archival observations. Mass profiles are derived by assuming hydrostatic equilibrium, and the dark matter distribution is fitted using the NFW profile.
In this study, we include only those systems for which the stellar masses are explicitly provided.

\subsection{Scatter in the observational data within the 3D parameter space}

Observationally estimated $V_{\mathrm{max}}$, $\bar{\Sigma}(<0.01r_{\mathrm{max}})$, $M_{\ast}$ for UFDs, classical dwarf spheroidal galaxies, massive galaxies, galaxy groups listed in the previous sections are confronted with theoretical predictions in Fig.~\ref{fig:fig3}.

Scatter along the $\bar{\Sigma}(<0.01\,r_{\mathrm{max}})$ axis (lower panels in Fig.~\ref{fig:fig3}):
In the $V_{\mathrm{max}}$–$\bar{\Sigma}(<0.01\,r_{\mathrm{max}})$ plane, noticeable scatter is seen in the $\bar{\Sigma}(<0.01\,r_{\mathrm{max}})$ direction for a fixed $V_{\mathrm{max}}$ (i.e., $M_{200}$). 
This dispersion implies diversity in the inner slopes of dark matter halo density profiles, suggesting that a binary classification (e.g., central slopes of $0$ or $-1$), as assumed in the theoretical models in this work, may be insufficient to describe the full range of observed structures.
Moreover, the same $V_{\mathrm{max}}$–$\bar{\Sigma}(<0.01\,r_{\mathrm{max}})$ projection reveals that, for fixed $M_{\ast}$, there is also considerable variation in $\bar{\Sigma}(<0.01\,r_{\mathrm{max}})$. 
This is likely the result of differences in internal halo structures and star formation efficiencies, which in turn affect the interplay between transition energy and available SNII energy. These findings support the interpretation that star formation efficiency is a key physical quantity in determining whether a cusp–core transition can occur via SNII feedback.

Scatter along the $V_{\mathrm{max}}$ axis (left panels in Fig.~\ref{fig:fig3}):
In the $V_{\mathrm{max}}$–$\bar{\Sigma}(<0.01\,r_{\mathrm{max}})$ plane, several galaxies exhibit lower values of $\bar{\Sigma}(<0.01\,r_{\max})$ than expected from the cusp plane, suggesting that cusp–core transitions have occurred in the past. These objects display significant scatter in $V_{\mathrm{max}}$, which implies a range of $M_{200}$ values resulting from variations in parameters such as the energy conversion efficiency $\varepsilon$ and the star formation efficiency.
Similarly, the $V_{\mathrm{max}}$–$M_{\ast}$ plane shows that, within the transition region, $M_{\ast}$ spans approximately an order of magnitude for a fixed $V_{\mathrm{max}}$ (or $M_{200}$), reinforcing the presence of diversity in star formation efficiencies. Since SNII energy scales with $M_{\ast}$, this variability leads to different outcomes with respect to cusp–core transformation, even under the assumption of a constant $\varepsilon$ for a given $M_{200}$.
These trends are also visible in the $M_{\ast}$–$\bar{\Sigma}(<0.01\,r_{\mathrm{max}})$ plane, which shows that dark matter haloes capable of undergoing a cusp–core transition span a wide mass range.

The lower-left panel of Fig.~\ref{fig:fig3} provides key insight into the connection between the theoretical expectations and observed halo structure. It complements Fig.~\ref{fig:fig2}, highlighting the features that distinguish haloes undergoing a cusp–core–cusp sequence, including the extent of structural transformation that occurs as haloes evolve through feedback-regulated processes.

%-----------
% Fig. 4
%-----------
\begin{figure}
    \centering
    \includegraphics[width=1.0\linewidth]{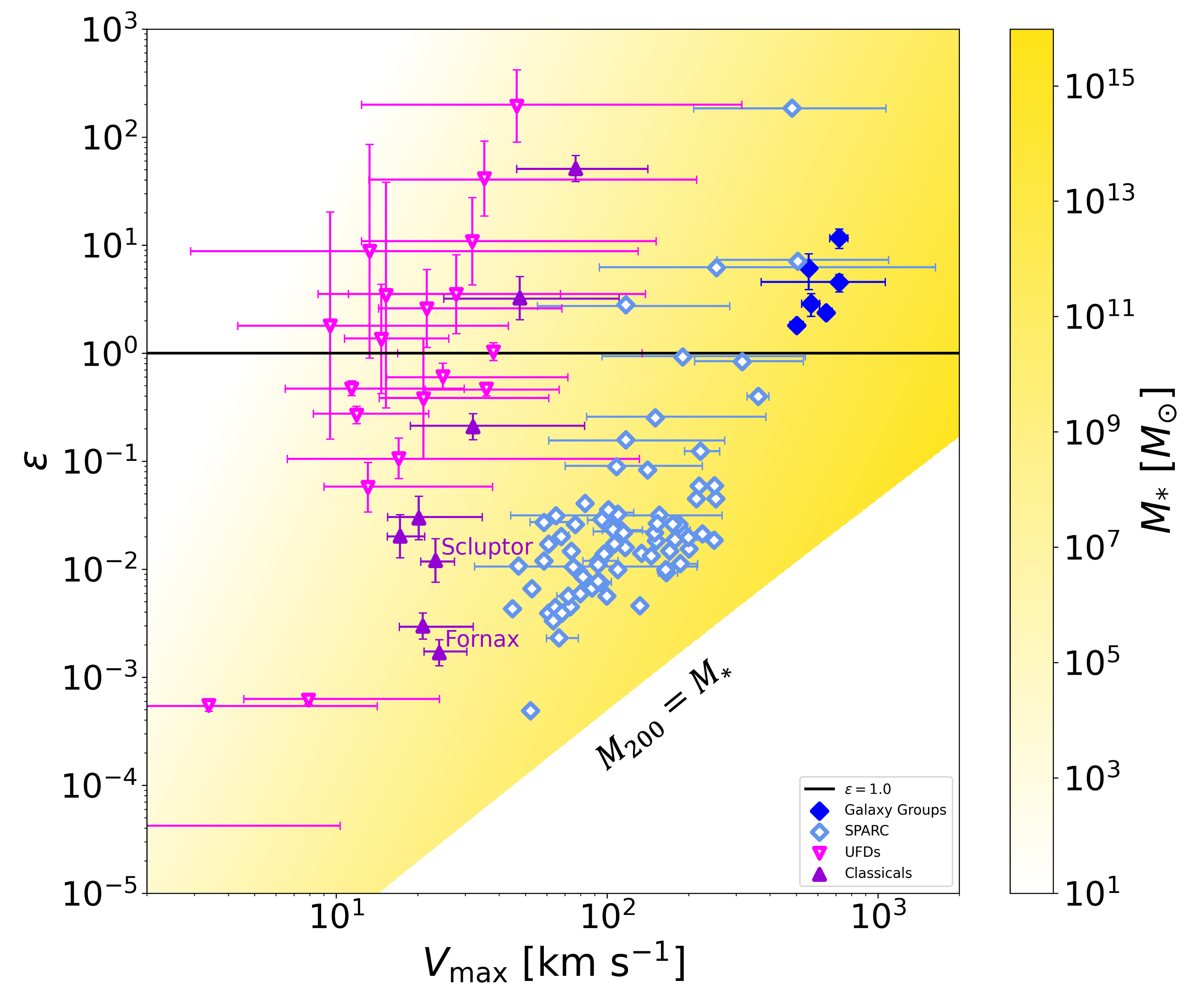}
    \caption{ 
    Relation between $V_{\max}$ and the energy conversion efficiency $\varepsilon$, derived from observed $V_{\max}$ and $M_{\ast}$. 
    %Symbols match those in the previous figure. 
    Background shading darkens with increasing stellar mass. The lower-right region, where stellar mass exceeds halo mass, is excluded. The solid line denotes $\varepsilon=1$ ($100\%$), above which cusp-core transitions are forbidden. SPARC data cluster around $\varepsilon\sim0.01$.
    }
    \label{fig:fig4}
\end{figure}

\subsection{Energy conversion efficiency from observational data}

Fig.~\ref{fig:fig4} shows the energy conversion efficiency $\varepsilon$ as a function of $V_{\mathrm{max}}$, derived from the observed stellar mass and maximum circular velocity using equation ~(\ref{M_star_crit}).
The stellar mass is indicated by the background colour map.
At a fixed $V_\mathrm{max}$, that is, at a fixed halo mass, $\varepsilon$ increases as the stellar mass decreases, implying that a lower supernova feedback energy results in a higher energy conversion efficiency.
The black solid line represents $\varepsilon=1$, corresponding to a $100\%$ conversion of available supernova feedback energy into gravitational potential energy.
Above this line lies a forbidden region in which cusp–core transitions are energetically forbidden because the available transition energy is insufficient.

The halo–mass–dependent trends visible in Fig.~\ref{fig:fig4} are consistent with those found in Fig.~\ref{fig:fig3}: galaxy groups lie within the forbidden region, whereas UFDs and classical dwarfs are scattered across the transition boundary.
\MS{
Regarding \MM{the} classical dwarf spheroidals, we highlight 
Fornax and Sculptor as well-studied benchmark systems with strong evidence for cored profiles. 
\MM{
The inferred energy conversion efficiencies is \MS{$1.20\%$ }for Sculptor and \MS{$0.17\%$ }for Fornax. 
}
Although the present analysis employs a two-phase cusp–core model, the results derived from these 
\MM{
well-characterised systems
}
allow us to extract physical significance beyond the limitations of the model.
}

In the upper-left panel of Fig.~\ref{fig:fig3}, SPARC galaxies are aligned parallel to the lines of constant energy conversion efficiency, depicted as solid and dashed lines, indicating a common trend in $\varepsilon$.
Based on the SPARC data, the characteristic efficiency $\varepsilon$ is found to be approximately $1\%$, suggesting that only about  $1\%$ of the available supernova feedback energy is sufficient to drive the cusp–core transition in these systems.
\MS{
Excluding unphysical 
\MM{
cases with
}
$\varepsilon>1$, the analysis of the SPARC data yields a mean of $\varepsilon=0.056$, and a median of $\varepsilon=0.017$. 
\MM{
Restricting the sample
}
to the main locus of the distribution ($10^{-3}<\varepsilon<10^{-1}$) results in a mean of $\varepsilon=0.020$ and a median of $\varepsilon=0.016$. 
}

\section{Discussion}
We now outline the implications for galaxy formation, considering the diversity of inner mass distributions, variation in star formation efficiency in low-mass galaxies, conditions for core stability, the complementary role of AGN feedback, the part played by energy-conversion efficiency in galaxy-formation modelling, and 
\MS{
the model limitations.
}

\subsection{Drivers of diversity in inner mass distributions}

The observed diversity in the inner dark matter mass distributions of dwarf galaxies arises naturally when the energy conversion efficiency between stellar feedback and the central gravitational potential is allowed to vary alongside the stellar-to-halo mass ratio. Empirical $z=0$ stellar-to-halo mass relations provide a useful baseline for this ratio \citep[e.g.][]{Moster+2013, Behroozi+2019}. At the extremely low stellar-to-halo mass ratios typical of UFDs, variations in star formation history, gas content, assembly pathways, and halo concentration at fixed virial mass can affect the balance between the energy required for a cusp–core transition and that provided by feedback. Conversely, in massive haloes such as galaxy groups and clusters, the deep potential well resists structural transformation.

These dependencies yield a spectrum of inner mass profiles at fixed $M_{200}$ and $V_{\mathrm{max}}$, ranging from persistent cusps to well-developed cores, along with corresponding spreads in characteristic surface densities. In Fig.~\ref{fig:fig3}, this variation manifests as substantial scatter in the characteristic mean surface density, particularly among UFDs. One possible contributor to this scatter is the difficulty of accurately constraining halo size for low-mass systems. Since $\varepsilon$ depends on both halo mass and stellar mass, uncertainties in either quantity can propagate into significant scatter in derived efficiency.
As evident in Fig.~\ref{fig:fig4}, UFDs show a substantial spread in $\varepsilon$ spanning $10^{-4}<\varepsilon<10^{2}$. This scatter indicates that variations in $\varepsilon$ likely drive the diversity of their inner mass distributions. Future work should investigate in detail the fundamental physical processes by which supernova feedback transfers energy to the gravitational potential of dark matter haloes.

In addition to observational uncertainty, physical processes governing star formation in dwarf galaxies are likely responsible for the diversity of stellar masses observed at fixed halo mass. 
\MM{
Several
}
principal mechanisms are known to suppress star formation in such systems. First, stellar feedback from supernova explosions can expel star-forming gas from shallow potential wells, effectively halting star formation \citep{DekelSilk1986,Mori+1997,ReadGilmore2005}
Second, the cosmic ultraviolet background, particularly during the epoch of reionisation, may prevent gas accretion and cooling in low-mass haloes. As reionisation proceeded earlier in overdense regions and later in underdense ones, the resulting spatial variation may have introduced environmental diversity in star formation histories \citep{DekelRees1987, Efstathiou1992,OkamotoFrenk2009}. 
Third, tidal interactions with a host halo can strip material from orbiting dwarf galaxies, further removing baryons and inhibiting star formation \citep{Mayer+2001, Fillingham+2015}. Fourth, due to their shallow gravitational potential wells, dwarf galaxies are particularly vulnerable to environmental influences such as ram pressure stripping caused by nearby massive galaxies \citep{MoriBurkert2000,Boselli+2022}.

\MM{
Recent work sharpens the picture. State-of-the-art zoom-in simulations show that ultraviolet background radiation, tides, and ram pressure in satellites can significantly suppress star formation at fixed halo mass \citep{Sawala+2016,Samuel+2022}. Within AGORA, the cross-code galaxy-simulation collaboration, \citet{Jung+2024} find that subhaloes with little or no starlight can form without tension with current observations, highlighting external gas removal and reionisation as key regulators of stellar mass at fixed halo mass. This connects directly to the present framework, in which external gas loss sets the available feedback energy and the energy conversion efficiency governs how strongly that energy couples to the central potential. Together, these effects explain the observed scatter and the diversity of central slopes as a joint dependence on the stellar-to-halo mass ratio and on coupling efficiency.
}

\subsection{Stability of the core}
While $M_{\ast}$ is one of the key quantities associated with the cusp–core transition, determining the relevant stellar mass is not straightforward. This is because stars formed after the transition do not contribute to the energy transfer involved in the process. Therefore, understanding the timescale over which the transition occurs is critical.
Because a change in the gravitational potential is required and the system must settle into a new equilibrium configuration, it may be more appropriate to consider the stellar mass formed within a dynamical time. This, however, raises the question of core stability. The present model assumes that the cored profile is inherently stable, without explicitly testing this assumption.

Although the precise timescale of the transition remains uncertain, \citet{OgiyaMori2011} showed that a single fluctuation in the potential is insufficient to stabilise the core; the system reverts to a cusp. Only when the potential is periodically-perturbed multiple times does the system settle into a stable core state \citep{OgiyaMori2014}. Given that star formation is expected to proceed cyclically in actual galaxies, periodic changes in potential are plausible. However, this implies that the transition must occur over a relatively long timescale, encompassing multiple episodes of star formation.

Moreover, \citet{Onorbe+2015} demonstrated that when the transition is triggered before the halo has fully relaxed into equilibrium, the core fails to remain stable and instead reverts to a cusp. Hence, the timing between the system's equilibration and the onset of potential fluctuations is also crucial. 
The present model implicitly assumes that the halo has reached equilibrium in its cuspy configuration before undergoing the transition.

\subsection{AGN feedback as a complementary channel}
Recent work has drawn attention to the possibility that active galactic nuclei (AGN) influence the inner structure of low–mass galaxies. Observational studies report elevated AGN incidence in dwarf galaxies, together with signatures of compact outflows and correlations with minor mergers in low–density environments, suggesting that nuclear activity may be more common and dynamically relevant at low masses than previously assumed \citet{MezcuaDomínguezSanchez2024} \citep[see also][]{Rodrguez+2025}. On the theoretical side, analytic arguments and cosmological simulations indicate that mechanically or radiatively driven AGN feedback can, under suitable conditions, produce central potential fluctuations and thereby assist cusp-core transformation \citep{Dashyan+2018, Koudmani+2025}.

An order–of–magnitude energetic estimate is given by
\begin{equation}
E_{\mathrm{AGN}} \simeq \varepsilon_\mathrm{r}\,\varepsilon_\mathrm{f}\,\Delta M_{\bullet} c^{2},
\end{equation}
where $\varepsilon_\mathrm{r}$ is the radiative efficiency, $\varepsilon_\mathrm{f}$ is the coupling fraction to the interstellar medium, and $\Delta M_{\bullet}$ is the accreted mass during an episode. Scaling to intermediate–mass black holes implies that $\Delta M_{\bullet}\sim 10^{4}$–$10^{5}\,M_{\odot}$ can yield $E_{\mathrm{AGN}}\sim 10^{56}$–$10^{57}\,\mathrm{erg}$ for fiducial efficiencies, comparable to or exceeding the supernova energy budget of dwarfs with $M_{\ast}\sim 10^{7}$–$10^{8}\,M_{\odot}$ (taking $E_{\mathrm{SN}}\sim 10^{56}$–$10^{57}\,\mathrm{erg}$ for a standard IMF). Thus, in systems that host nuclear black holes, AGN episodes could in principle compete with or supplement supernova feedback in driving central potential fluctuations.

Several observational discriminants follow if AGN contribute to cusp to core evolution. One expects nuclear outflows with velocities of several $100~\mathrm{km\,s}^{-1}$ in optical and near–infrared lines, compact radio cores or jets aligned with disturbed central gas morphologies, spatial coincidences between outflow cavities and deficits in molecular or atomic gas mapped by ALMA or high–resolution H\,{\sc i}, and correlations between indicators of recent AGN variability and unusually shallow inner mass profiles at fixed $M_{\ast}$ and star formation rate. These predictions are testable with current integral-field spectroscopy and JWST observations on kiloparsec and sub-kiloparsec scales.

Important caveats remain. The black hole occupation fraction in dwarf galaxies is uncertain, AGN samples may be contaminated by $X$-ray binaries, duty cycles and variability are poorly constrained, and the coupling efficiency $\varepsilon_\mathrm{f}$ in multiphase gas is not well known. Moreover, energy deposition that proceeds adiabatically or predominantly at large radii will not efficiently modify the collisionless component. These effects likely introduce substantial galaxy-to-galaxy scatter in any AGN-driven contribution to core formation.
Given these uncertainties and the strong empirical support for stellar feedback at low masses, we focus here on supernova-driven coupling and quantify its connection to inner structure through the energy conversion efficiency. The AGN channel is treated as a plausible, complementary pathway that can be incorporated in future work by extending the same energetic framework to include an explicit AGN term in the coupling budget.

\subsection{The role of energy conversion efficiency in modelling of galaxy formation}
We expect that the influence of energy conversion efficiency extends beyond the cusp--core problem to broader issues in cosmological structure formation and galaxy formation. Defined as the fraction of available supernova energy that couples to the dark matter potential, $\varepsilon$ is a model independent quantity that does not hinge on a particular code architecture, resolution strategy, or initial setup. In this sense it can serve as a shared reference for comparing how different stellar feedback prescriptions transfer energy to haloes and regulate the baryon cycle from the interstellar medium to the circumgalactic medium.

A wide variety of supernova feedback models are in active use. Examples include kinetic wind prescriptions that scale with halo properties in large volume runs such as IllustrisTNG \citep{Pillepich+2018,Nelson+2019}, thermal stochastic injection as in EAGLE \citep{DallaVecchiaSchaye2012,Schaye+2015,Crain+2015}, explicitly multi channel treatments of radiation, winds, and supernovae as in FIRE and FIRE 2 \citep{Hopkins+2018}, blastwave and delayed cooling schemes
\MM{
\citep{Mori+1997} 
}
with early stellar feedback extensions as in MaGICC and NIHAO \citep{Stinson+2006,Stinson+2013,Wang+2015}, superbubble models in GASOLINE2 \citep{Keller+2014,Keller+2015}, mechanical feedback that injects terminal momentum in RAMSES based implementations \citep{KimmCen2014}, momentum driven formulations that resolve the snowplow scaling \citep{KimOstriker2015}, and modular winds within frameworks such as MUFASA, and  SWIFT\citep{Dave+2016,Schaller+2024}. 

Against this landscape, our observationally inferred value, $\varepsilon \approx 0.01$, suggests a gentle calibration target that can be applied across these diverse prescriptions. This estimate should be refined with larger samples and across a range of environments, but using $\varepsilon$ as a reference encourages feedback schemes to reproduce a common level of energy coupling when averaged over realistic burst cycles and geometries. Doing so can narrow freedom in mass loading, momentum per supernova, coupling timescales, and the partition of energy among thermal, kinetic, radiative, and nonthermal channels. With such a reference in place, population level predictions can become less sensitive to the choice of feedback model, including the stellar to halo mass relation, disk sizes and angular momentum, gas fractions and metallicity gradients, and the demographics of dwarf and satellite systems. In this view, local energy transfer is linked to global statistics without prescribing a unique implementation. In practice, simulations would be evaluated not only by matching internal kinematics, but also by remaining consistent with the observed range of $\varepsilon$ while reproducing the distribution of galaxy properties across mass and environment.

The same reference can inform semi analytic models and related frameworks. Implementing $\varepsilon$ as a mild constraint translates supernova energy budgets into a physically anchored adjustment of halo structure, in place of ad hoc coupling factors. In application, $\varepsilon$ can guide the calibration of mass loading, duty cycles, outflow escape fractions, and time dependent feedback kernels that regulate gas supply, star formation, and metal enrichment. It can also bound predicted core sizes and inner slopes at fixed halo mass, improve the mapping between star formation histories and halo response, and indicate when additional physics may be required if $\varepsilon$ compatible feedback does not suffice. In this way, an observation based reference for $\varepsilon$ can propagate through forecasts for scaling relations, the connection between haloes and galaxies, and the demographics of cores, while keeping the interpretation of galaxy formation models as transparent as possible.

\MM{
\subsection{Applicability and future directions}
}

\MM{
Here we discuss dynamical extensions and effects not treated in the present model, spanning both inner and outer halo structure, with the aim of clarifying where the current framework applies and how it can be strengthened. 
}

\MM{
On the internal side, the two structural phases used here (an NFW, cusped phase and a Burkert, cored phase) can be generalised to a continuous family of central profiles, allowing the inner slope to vary smoothly and to match the observed continuum of central structures. In practice, this entails adopting flexible density laws such as an Einasto profile \citep{Einasto1965} or a generalised Hernquist profile \citep{Hernquist1990}, estimating the inner slope jointly with scale radii, and thereby enabling a more precise characterisation of the transition that accommodates the diversity seen in observations.
}
\MM{
On the external side, environment-driven evolution should be incorporated. Tidal forces from a host halo can truncate the outer profile, alter the virial radius, and change the mapping between $M_{200}$, $V_\mathrm{max}$, and $\bar{\Sigma}(<0.01\,r_\mathrm{max})$. This necessitates relaxing the fixed-$r_{200}$ assumption and tracking the coupled evolution of mass, radius, and concentration during stripping.
}

\MM{
Building on the foregoing, a natural next step is to relax the assumption of instantaneous star formation and to examine possible redshift dependence, thereby enabling a more detailed, observationally anchored assessment of how feedback couples to halo structure across cosmic time.
}\bigskip

\section{Conclusions}
We present an energetics-based model of the cusp-core transition tied to the $c$--$M$ relation, quantify the required transition energy and the associated transition-forbidden region, and infer from SPARC data an energy-conversion efficiency. The key points are:

\begin{itemize}

\item We expressed the total energy of dark matter haloes before and after the transition as a function of the one-parameter family defined by the virial mass, $M_{200}$, assuming the $c$--$M$ relation. The gravitational potential energy for the NFW profile is obtained analytically, while that for the Burkert profile is approximated using a calibrated fitting function. Both $\psi_{\mathrm{NFW}}$ and $\psi_{\mathrm{BKT}}$ vary only weakly with $M_{200}$.

\item By computing the associated energy difference $\Delta E$ across $M_{200}=10^{5}$–$10^{16}\,M_{\odot}$, we find that $\Delta E$ is typically $\simeq 10\%$ of the total gravitational energy of the initial NFW halo. The occurrence of the transition can therefore be framed as whether this $\simeq 10\%$ threshold is met.

\item We assessed whether SNII feedback, taken as the dominant source, can supply the transition energy. Using a Chabrier IMF, we estimated the SNII energy per unit stellar mass and thus derived the critical stellar mass $M_{\ast,\mathrm{crit}}$ required to trigger the cusp-core transition.

\item Incorporating the $c$--$M$ relation, we examined the interplay between halo virial mass, stellar mass and characteristic mean surface density in a three-dimensional parameter space. This analysis identifies a transition-forbidden region in which SNII feedback is energetically unable to induce the transformation.

\item We tested the model against observational data spanning dwarf galaxies to galaxy clusters, after mapping theoretical quantities to observables. Most galaxy-scale systems lie outside the transition-forbidden region, whereas most galaxy clusters lie within it.

\item The spread in the characteristic mean surface density of dwarf-scale haloes, which reflects variations in their central density slopes, indicates that their profiles are not limited to pure NFW or Burkert forms. This diversity is naturally explained by variations in star formation efficiency and energy conversion efficiency, reinforcing their roles in governing the cusp-core transition.

\item Combining observations with our cusp-core transition model, we infer the energy-conversion efficiency $\varepsilon$: galaxy groups lie in the transition-forbidden regime with $\varepsilon>1$, dwarfs scatter around the $\varepsilon=1$ boundary, and SPARC galaxies show a roughly lognormal $\varepsilon$ with a median of a few per cent, indicating that only a small fraction of supernova feedback energy is needed to drive the cusp-core transition in typical disc galaxies.

\item Beyond the cusp-core problem, the energy-conversion efficiency $\varepsilon$ (the fraction of supernova energy coupled to the dark-matter potential) provides a model-independent calibration. An observationally inferred $\varepsilon\approx 0.01$ offers a common target for tuning feedback prescriptions and semi-analytic models.

\end{itemize}

%--------------------------------------------------------------------------------------
%\clearpage
%--------------------------------------------------------------------------------------

\section*{Acknowledgements}
We thank the anonymous referee for helpful comments.
We are grateful to Go Ogiya, Hidenobu Yajima, Alexander Y. Wagner, Giulia Despali and Jinning Liang for their valuable comments.
This work was supported by JSPS KAKENHI Grant Numbers JP23KJ0280~(Y.K.), JP24K07085~(M.M.), JP25H01553 (K.H.) and JP24K00669~(M.M., K.H.) and by NINS Astrobiology Center program research (Grant Number API07067).
Numerical computations were performed with computational resources provided by the Multidisciplinary Cooperative Research Program in the centre for Computational Sciences, University of Tsukuba.

%%%%%%%%%%%%%%%%%%%%%%%%%%%%%%%%%%%%%%%%%%%%%%%%%%
\section*{Data Availability}

Data related to this work will be shared on reasonable request to the corresponding author.



\bibliographystyle{mnras}
\bibliography{ref} % if your bibtex file is called example.bib

%\begin{thebibliography}{99}
%\end{thebibliography}

%%%%%%%%%%%%%%%%%%%%%%%%%%%%%%%%%%%%%%%%%%%%%%%%%%

%%%%%%%%%%%%%%%%% APPENDICES %%%%%%%%%%%%%%%%%%%%%

%\appendix


%%%%%%%%%%%%%%%%%%%%%%%%%%%%%%%%%%%%%%%%%%%%%%%%%%

% Don't change these lines
\bsp	% typesetting comment
\label{lastpage}
\end{document}